# Identification time-delayed fractional order chaos with functional extrema model via differential evolution

Fei Gao[a,b,*], Xue-jing Lee[a], Feng-xia Fei[a], Heng-qing Tong[a]

[a]*Department of Mathematics, School of Science, Wuhan University of Technology, Luoshi Road 122, Wuhan, Hubei,430070, People's Republic of China*
[b]*Signal Processing Group, Department of Electronics and Telecommunications, Norwegian University of Science and Technology, N-7491 Trondheim, Norway*

**Abstract**

In this paper, a novel inversion mechanism of functional extrema model via the differential evolution algorithms(DE), is proposed to exactly identify time-delays fractional order chaos systems. With the functional extrema model, the unknown time-delays, systematic parameters and fractional-orders of the fractional chaos, are converted into independent variables of a non-negative multiple modal functions' minimization, as a particular case of the

✩The work is partially supported by Supported by the NSFC projects No.81271513 of China, the Fundamental Research Funds for the Central Universities of China, the self–determined and innovative research funds of WUT No. 2012–Ia–035, 2013-Ia-040, 2012-Ia-041, 2010–Ia–004, Scientific Research Foundation for Returned Scholars from Ministry of Education of China(No. 20111j0032), the HOME Program No. 11044(Help Our Motherland through Elite Intellectual Resources from Overseas) funded by China Association for Science and Technology, the Natural Science Foundation No.2009CBD213 of Hubei Province of China, The National Soft Science Research Program 2009GXS1D012 of China, the National Science Foundation for Post–doctoral Scientists of China No. 20080431004. The work was carried out during the tenure of the ERCIM Alain Bensoussan Fellowship Programme, which is supported by the Marie Curie Co-funding of Regional, National and International Programmes (COFUND) of the European Commission.

*Corresponding author

*Email addresses:* hgaofei@gmail.com (Fei Gao), 695459581@qq.com (Xue-jing Lee), 1092285218@qq.com (Feng-xia Fei), tonghq2005@whut.edu.cn (Heng-qing Tong )

*URL:* http://feigao.weebly.com (Fei Gao)

functional extrema model's minimization. And the objective of the model is to find their optimal combinations by DE in the predefined intervals, such that the objective functional is minimized. Simulations are done to identify two classical time-delayed fractional chaos, Logistic and Chen system, both in cases with noise and without. The experiments' results show that the proposed inversion mechanism for time-delay fractional-order chaotic systems is a successful methods, with the advantages of high precision and robustness.

## 1. Introduction

Recently, the topic of fractional-order time-delayed differential equation has attracted growing interest among Mathematicians and Physicists [1–4]. Specially, fractional-order delayed differential systems can characterize chaotic behaviors [1–6].

The applications of fractional differential equations, whose nonlinear dynamics are described by a powerful tool with the concept of fractional calculus[7–12], began to appeal to related scientists[4, 13–24] in following areas, bifurcation, hyperchaos, proper and improper fractional-order chaos systems and chaos synchronization[14–17, 21–29].

Fractional delayed differential equation (FDDE) is a differential equation in which the fractional derivative of the function at any time depends on the solution at previous time[1, 4–6]. Introduction of delays in the model enhances its dynamics and allows a precise description of real life phenomena. In FDDE, history of the system over the delay interval $[\tau, 0]$ is provided as the initial condition. Many synchronization methods are valid for fractional-order chaotic systems with known parameters and time–delays[30–34]. However, in some applications such as secure communications and chaos synchronization, fractional chaotic system is partly known. That is, the form of differential equation is known, however, some or all of the time-delays, fractional orders and parameters are unknown. Then estimating the unknown parameters of the chaos are really important in controlling and utilizing chaos, for the fractional chaos both with time-delay and without. Thus, it is of vital importance to identify the unknown time-delays, fractional orders and parameters in fractional-order chaotic delayed systems. And it is difficult to identify the and time-delays and parameters in the uncertain fractional-order chaotic systems[4, 4, 16, 17, 35, 36], as well as the normal unknown delayed dynamical complex networks or hyper chaos[37, 38].

The process to get the exact values of uncertain orders and parameters for the fractional order chaotic systems is called system inversion mechanism. Hitherto, there have been some approaches in system inversion for fractional-order chaos systems. For instances, synchronization for fractional-order chaos[35] and fractional order complex networks[36]. However, the design of controller and the updating law of parameter identification is still a hard task with techniques and sensitivities depending on the considered systems. And the non-classical way via artificial intelligence methods, for examples, differential evolution[16] and particle swarm optimization [4, 17]. In which only the commensurate fractional order with the same $q$ chaos sys-tems and simplest case with one unknown fractional order $q$ for fractional-order chaos systems are discussed. But no studies on some of the unknown time-delays $\tau_1, \tau_2, ..., \tau_n$ are done up till now. In recent results[4], only the cases are discussed in unknown partial parameters $(\theta_1, \theta_2, ..., \theta_m)$, fractional orders $(q_1, q_2, ..., q_k)$ but known time-delays $(\tau_1, \tau_2, ..., \tau_n)$. And in reference [39], only the identification for fractional chaos system without time-delays are discussed.

However, to the best of our knowledge, few work in non-classical way has been done to the parameters, time-delays and fractional orders $(\theta_1, \theta_2, ..., \theta_m, \tau_1, \tau_2, ..., \tau_n, q_1, q_2, ..., q_k)'$ together inversion estimation for time-delayed fractional-order chaos systems so far.

The objective of this work is to present a novel simple but effective inversion mechanism with a functional extrema model of uncertain fractional-orders and parameters based on differential evolution algorithms (DE) to estimate the time-delayed fractional order chaotic systems. In which, unknown time-delays, fractional-orders and parameters are estimated together in a non-Lyapunov way. And the illustrative inversion simulations in different chaos systems system are discussed respectively.

The rest is organized as follows. Section 1 give a simple review on parameters inversion mechanism for time-delayed fractional-order chaos systems. In section 2, firstly the DE are introduced briefly. Then a novel inversion methods with functional extrema minimization model based on DE is proposed to estimate the fractional chaos systems. In section 3, simulations are done to two classical time-delayed fractional order chaotic systems. And results analysis and discussions are done too. Conclusions are summarized briefly in Section 4.

## 2. Inversion mechanism with functional extrema model

*2.1. Inversion mechanism with functional extrema model via DE*

We consider the following fractional-order chaos system with time-delays.

$$_\alpha \mathscr{D}_t^q Y(t) = f\left(Y(t), Y(t-\tau), Y_0(t), \theta\right) \quad (1)$$

The continuous integro-differential operator[40, 41] is defined as

$$_\alpha \mathscr{D} y_t^q = \begin{cases} \frac{d^q}{dx^q}, q > 0; \\ 1, q = 0; \\ \int_\alpha^1 (d\tau)^q. \end{cases}$$

Let $Y(t) = (y_1(t), ..., y_n(t))^T \in \Re^n$ denotes the state vector. $\theta = (\theta_1, \theta_2, ..., \theta_n)^T$ denotes the original parameters. $q = (q_1, q_2, ..., q_n), (0 < q_i < 1, i = 1, 2, ..., n)$ is the fractional derivative orders.

Normally the function $f = (f_1, f_2, ..., f_n)$ is known. And the $\theta = (\theta_1, \theta_2, ..., \theta_n)$, $q = (q_1, q_2, ..., q_n)$ will be the parameters to be estimated. Then a correspondent system are constructed as following.

$$_\alpha \mathscr{D}_t^{\tilde{q}} \tilde{Y}(t) = f\left(\tilde{Y}(t), \tilde{Y}(t-\tilde{\tau}), Y_0(t), \tilde{\theta}\right) \quad (2)$$

where $\tilde{Y}(t), \tilde{\tau}, \tilde{\theta}, \tilde{q}$ are the correspondent variables to those in equation (1), and function $f$ are the same. The two systems (1) (2) have the same initial condition $Y_0(t)$.

Let $L(t) = (Y_1(t), Y_2(t), ..., Y_K(t))$ and $L(t) = (\tilde{Y}_1(t), \tilde{Y}_2(t), ..., \tilde{Y}_K(t))$, where $1, 2, ..., K$ is the sampling time point and $K$ denotes the length of data used for identification. $\|\cdot\|$ is Euclid norm.

Then the objective is obtained as following,

$$(\theta, q, \tau)^* = \underset{(\theta,q,\tau)}{\arg\min} F = \underset{(\theta,q,\tau)}{\arg\min} \sum_t \left\| L(t) - \tilde{L}(t) \right\|_2 \qquad (3)$$

When some the fractional chaotic differential equations $f = (f_1, f_2, ..., f_n)$ are unknown, the objective will be,

$$(f_1, f_2, ..., f_n)^* = \underset{(f_1, f_2, ..., f_n)}{\arg\min} F \qquad (4)$$

Equation (4) is fractional-order chaos' inversion mechanism called reconstruction. In Reference [42] a novel method was proposed to reconstruct the unknown equations $(f_1, f_2, ..., f_n)$ based on an united mathematical model. However, for the united mathematical model[42], to be identified is only $(f_1, f_2, ..., f_n)$ instead of $q$. That is, the left part $_\alpha\mathscr{D}_t^q \tilde{Y}(t)$ of the equation (2) are not included.

Therefore, to estimate the $q$ of the equation (2) with unknown systematic parameters $\theta$ is still a question to be solved for parameters and orders estimation of non-commensurate and hyper fractional-order chaos systems.

It can be derived into a functional extrema model as following equation (5). In which, the two different Equation (3) and Equation (4) are two different aspects. One is to find best the parameters as in Equation (3), and the other is to find best the functionals to obtain the minimums as in Equation (4).

$$J(f_1, ..., f_n) = \sum_t \left\| L(t) - \tilde{L}(t) \right\|_2 \qquad (5)$$

Equation (5) is the functional extrema model we proposed for time-delayed fractional order systems. Then a novel inversion mechanism by Dif-ferential evolution algorithm (DE) as following Figure 1.

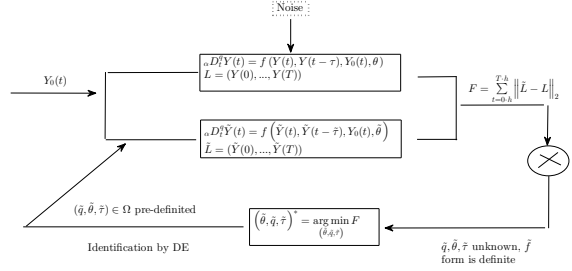

Figure 1: Novel inversion mechanism by Differential evolution algorithm (DE)

Now we take the time-delayed fractional order Logistic system (6)[40, 43] for instance.

$$\begin{aligned} _0\mathscr{D}_t^q x(t) &= -ax(t) + \gamma x(t-\tau) - \gamma x^2(t-\tau) \\ x(t) &= 0.5, \quad t \in [-\tau, 0] \end{aligned} \quad (6)$$

when $q = 0.9, a = 26, \tau = 0.5, \gamma = -53$, system (6) is chaotic. And Figure 2 shows the chaotic of Logistic system (6).

Secondly the objective function for time-delay system (6) is chosen as:

$$J = F(q, a, \tau, \gamma) = \sum_{t=0 \cdot h}^{T \cdot h} \left\| L(t) - \tilde{L}(t) \right\|_2 \quad (7)$$

Equation (7) is a special case of functional extrema model Equation (5), with the particular independent variables of $q, a, \tau, \gamma$. And the objective function for system (6) is shown as Figure 3.

From the Figure 3, we can see that the best combination, of the independent variables of $q, a, \tau, \gamma$, is a global minimum of the objective function (7), which is a multi-modal function with a lot of local minimum. Then we can conclude that the general case of the function (7), that is functional extrema model (5), is also multi-modal.

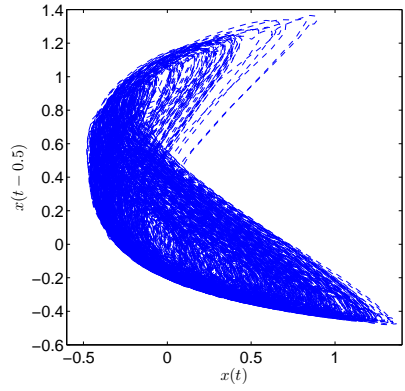

Figure 2: Logistic system (6)

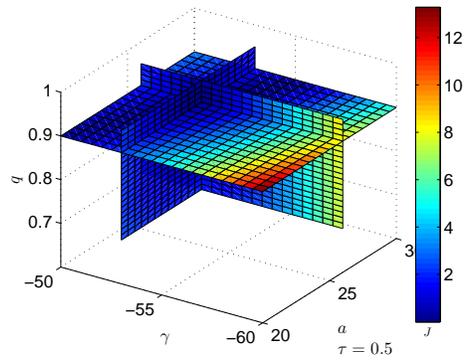

Figure 3: The objective function of time-delayed fractional order Logistic system

Actually, the inversion mechanism for fractional chaos is to find the best form of equations for functional extrema model (5). In this paper, the task is to find the best combination of the independent variables of $q, a, \tau, \gamma$ for the objective function (7), as a particular case.

*2.2. A simple review of Differential evolution (DE) algorithm*

Differential Evolution (DE) algorithm grew out of Price's attempts to solve the Chebychev Polynomial fitting Problem that had been posed to him by Storn [44]. A breakthrough happened, when Ken came up with the idea of using vector differences for perturbing the vector population. Since this seminal idea, a lively discussion between Ken and Rainer and endless ruminations and computer simulations on both parts yielded many substantial improvements which make DE the versatile and robust tool it is today [44–47].

DE utilizes $M$ $n$–dimensional vectors, $X_i = (x_{i1}, \cdots, x_{in}) \in S, i = 1, \cdots, M$, as a population for each iteration, called a generation, of the algorithm. For each vector $X_i^{(G)} = (X_{i\,1}^{(G)}, X_{i\,2}^{(G)}, \cdots, X_{i\,n}^{(G)}), i = 1, 2, \cdots, M$, there are three main genetic operator acting.

For each individual, to apply the mutation operator, firstly random choose four mutually different individual in the current population $X_{r_1}^{(G)}, X_{r_2}^{(G)}, X_{r_3}^{(G)} (r_1 \neq r_2 \neq r_3 \neq i)$. Then combines it with the current best individual $X_{best}^{(G)}$ to get a perturbed vector $V = (V_1, V_2, \cdots, V_n)$ [44, 48] as below:

$$V = \begin{cases} X_{r_3}^{(G)} + 0.5(CF+1) \cdot \left(X_{r_1}^{(G)} + X_{r_2}^{(G)} - 2X_{r_3}^{(G)}\right), if \ rand(0,1) < 0.5 \\ X_{r_3}^{(G)} + CF \cdot \left(X_{r_1}^{(G)} - X_{r_2}^{(G)}\right), otherwise \end{cases} \tag{8}$$

where $CF > 0$ is a user-defined real parameter, called mutation constant, which controls the amplification of the difference between two individuals to avoid search stagnation.

Following the crossover phase, the crossover operator is applied on $X_i^{(G)}$. Then a trial vector $U = (U_1, U_2, \cdots, U_n)$ is generated by:

$$U_m = \begin{cases} V_m, & if \ (rand(0,1) < CR) \ or \ (m = k), \\ X_{i\,m}^{(G)}, & if \ (rand(0,1) \geq CR) \ and \ (m \neq k). \end{cases} \tag{9}$$

in the current population[44], where $m = 1, 2, \cdots, n$, the index $k \in \{1, 2, \cdots, n\}$ is randomly chosen, $CR$ is a user-defined crossover constant[44, 48] in the range $[0, 1]$. In other words, the trial vector consists of some of the components of the mutant vector, and at least one of the components of a randomly selected individual of the population.

Then it comes to the replacement phase. To maintain the population size, we have to compare the fitness of $U$ and $X_i^{(G)}$, then choose the better:

$$X_i^{(G+1)} = \begin{cases} U, if \ F(U) < F(X_i^{(G)}), \\ X_i^{(G)}, otherwise. \end{cases} \tag{10}$$

The pseudo-code of the DE is given below as Algorithm 1.

**Algorithm 1** A novel inversion mechanism based on differential evolution algorithms
---
1: **Basic parameters' setting for DE**.
2: **Initialize** Generate the initial population.
3: **while** Termination condition is not satisfied **do**
4:     **Evaluation** Evaluate the fitness and remain the best individual.
5:     **Mutation** As in equation (8).
6:     **Crossover** As in equation (9).
7:     **Replacement** As in equation (10).
8:     **Boundary constraints** For each $x_{ik} \in X_i, k = 1, 2, ..., D$, if $x_{i1}$ is beyond the boundary, it is replaced by a random number in the boundary.
9: **end while**
10: **Output** Global optimum $x_{Best}$

## 3. Simulations and discussion

*3.1. Simulations to the time-delayed fractional chaotic systems*

We take two classical time-delayed fractional chaotic systems with cases with noise and without noise. One is the time-delay Logistic system (6). And another is the fractional order time-delayed Chen system (11)[3, 4] as following.

$$\begin{cases} {}_0\mathscr{D}_t^q x(t) = a(y(t) - x(t-\tau)); \\ {}_0\mathscr{D}_t^q y(t) = (c-a)x(t-\tau) - x(t)z(t) + cy(t); \\ {}_0\mathscr{D}_t^q z(t) = x(t)y(t) - bz(t-\tau). \end{cases} \quad (11)$$

And when $(a,b,c) = (35,3,27)$, $q = 0.97, \tau = 0.005$, $x(t) = 0.2, y(t) = 0, z(t) = 0.5$, $t \in [-\tau, 0]$, Chen system (11) is a chaotic system[3]. And Figure 4 shows the chaotic of Chen system (11).

And for Chen system (11), the independent variables of $q, a, b, c, \tau$. The objective function for system (11) is shown as Figure 5.

For the the time-delay Logistic system (6), step $h = 0.01$, unknown independent variables $a, \gamma, \tau, q$ are predefined in $\Omega = [20, 40] \times [-60, -50] \times [0.1, 2.51] \times [0.1, 0.99]$, and to be estimated. For the fractional order time-delayed Chen system (11), step $h = 0.001$, unknown independent variables $a, b, c, \tau, q$ are predefined in $\Omega = [30, 40] \times [0.001, 10] \times [20, 30] \times [0.001, 0.009] \times [0.001, 1]$ and are to be identified.

For the solving methods for fractional time-delayed system, the method[3, 4] are selected. It is originated from a modification of AdamsBashforthMoulton algorithm (Predictorcorrector approach) is proposed by Diethelm[49] to solve fractional differential equations. And it was extended for fractional time-delayed differential equations[50]. And to resolve the random fractional systems with delays generated in evolution process of differential evolution algorithms , the No. of samples are chosen 100 for the methods[3, 4, 50].

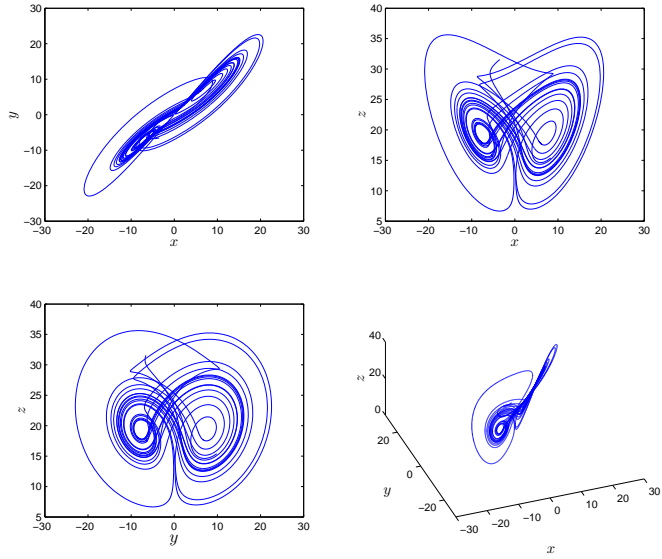

Figure 4: Chen system (11)

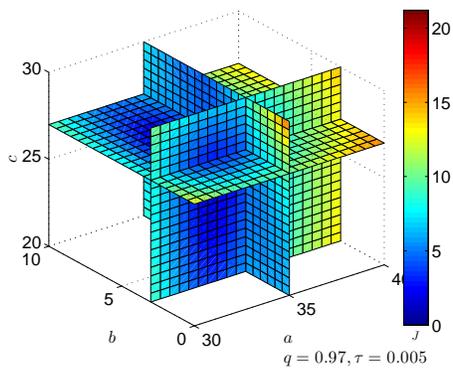

Figure 5: The objective function of time-delayed fractional order Chen system

And numerical simulations are done with the measured response $x^m(t_j)$ is obtained by adding noise, say $x^m(t_j) = (1 \pm NSR * r_j)\tilde{x}^m(t_j)$, where NSR is the noise-to-signal ratio (i.e., between 0.1% and 3%), $r_j$ is a random number with uniform distribution within the interval $[0, 1]$ (note: a different random number for each sample), and $\tilde{x}^m(t_j)$ is the free-noise response of the original systems.

For systems to be identified, the parameters of the proposed method are set as following. The parameters of the simulations are fixed: the size of the population was set equal to $M = 40$, generation is set to 300, the default values $CF = 1$, $CR = 0.85$; The times of function evaluations are 10020. Table 1 shows the simulation results of above fractional order chaotic systems.

As table 1 showed, it can concluded that if we add noise to the signal, then it is really not easy to achieve the genuine values of time-delays and systematic parameters.

Figure 6 shows the evolution process of DE for time-delayed Logistic system from one simulations.

Figure 7 shows objective and correspondent time-delay and parameters for Logistic system from simulations.

Figure 8 shows objective and correspondent time-delay and parameters for Chen system from simulations.

Figure 9 shows objective functions' value of all the simulations for Chen

Table 1: Simulation results for time delayed fractional order chaos systems (TDFOC)

| TDFOC | NSR | StD | Mean | Min | Max | Success rate[a] |
|---|---|---|---|---|---|---|
| Logistic | 0 | 3.3393e-06 | 5.3457e-07 | 1.8707e-09 | 3.3384e-05 | 99% |
| Logistic | ±0.005 | 0.001354 | 0.020187 | 0.017381 | 0.023116 | 0% |
| Logistic | ±0.01 | 0.0024677 | 0.03993 | 0.034891 | 0.047666 | 0% |
| Logistic | ±0.015 | 0.0044955 | 0.059172 | 0.047256 | 0.067368 | 0% |
| Logistic | ±0.03 | 0.0071113 | 0.12006 | 0.10006 | 0.13829 | 0% |
| Chen | 0 | 6.9667e-07 | 1.0968e-06 | 8.5626e-08 | 2.9771e-06 | 100% |
| Chen | ±0.005 | 0.026409 | 0.9144 | 0.86426 | 0.95854 | 0% |

[a] Success means the the solution is less than $1e-5$ in 100 independent simulations.

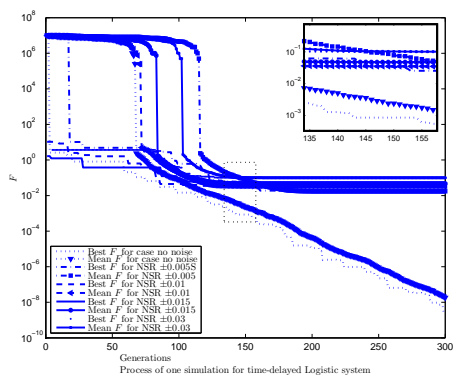

Figure 6: Evolution process of DE for time-delayed Logistic system from simulations

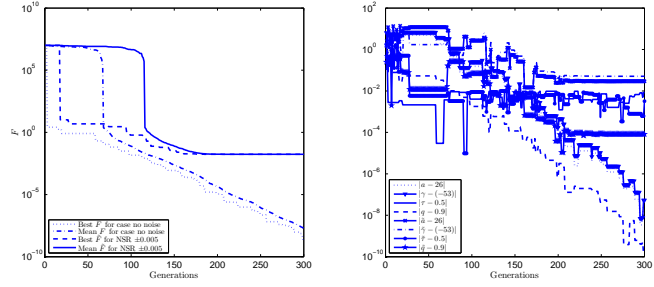

Figure 7: The objective and correspondent time-delay and parameters for Logistic system

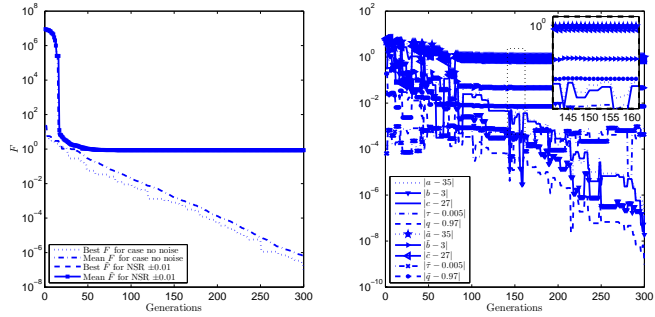

Figure 8: The objective and correspondent time-delay and parameters for Chen system

and Logistic systems.

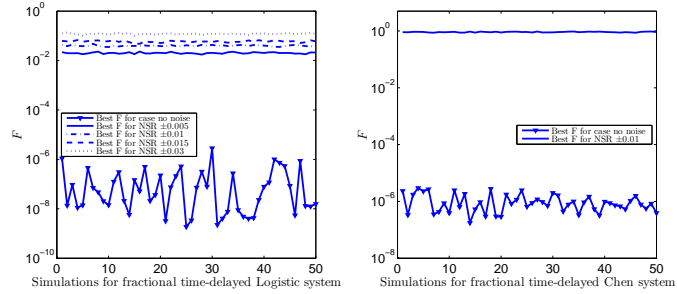

Figure 9: The objective functions' value for all the simulations of Chen and Logistic systems

When $q = 0.94, \alpha = 0.009$, system (11) also shows chaotic[3]. And Figure 10 show the objective value and correspondent time-delay and parameters for Chen system with different $q, \alpha$. Here unknown independent variables $a, b, c, \tau, q$ are predefined in $\Omega = [30, 40] \times [0.001, 10] \times [20, 30] \times [0.001, 0.03] \times [0.001, 1]$.

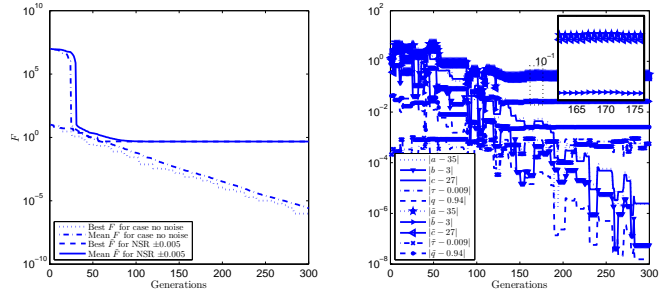

Figure 10: The objective and correspondent time-delay and parameters for Chen system with different $q, \alpha$

As the above figures showed, the DE is effective for identification the unknown time-delays and parameters of time-delayed Logistic and Chen chaotic system.

*3.2. Results analysis and discussions*

From the simulations results above, it can be concluded that the proposed method is efficient.

To test the performance of the proposed method Algorithms 1 , some more simulations are done to time delayed fractional order Chen systems in following cases A,B,C,D. In these cases, each with only one condition is changed according to the original setting for Chen system.

- Case A. Enhancing the predefined intervals from $\Omega = [30, 40] \times [0.001, 10] \times [20, 30] \times [0.001, 0.009] \times [0.001, 1]$ to $\Omega = [30, 40] \times [0.001, 10] \times [20, 30] \times [0.001, 0.3] \times [0.001, 1]$.

- Case B. Minimizing the number of samples for resolving system (11) from 100 to 50.

- Case C. Changing the iteration numbers of DE from 300 to 600.

- Case D. Changing the population size of DE from 40 to 80.

And the simulation results are listed in Table 2.

From results of the Table 2, we can conclude that minimizing the number of samples for computing the system (11) as case B, enhancing the generation numbers of DE as case C, the population size of Algorithms 1 as case D, will make the Algorithms 1 much more efficient and achieve a much more higher precision. However if the predefined intervals of the system (11) are enhanced, then the results will go to the opposite way. That is the success rate is from 99% to 18% as case A.

Table 2: Simulation results for system (11)

| system (11) | StD | Mean | Min | Max | Success rate[a] | NEOF[b] |
|---|---|---|---|---|---|---|
| Case A. | 1.8690e-01 | 1.3022e-01 | 5.7984e-05 | 3.4439e-01 | 18% | 12040 |
| Case B. | 1.6286e-07 | 2.6430e-07 | 7.6571e-08 | 3.6769e-07 | 100% | 12040 |
| Case C. | 1.6618e-13 | 5.1275e-13 | 3.4319e-13 | 6.7534e-13 | 100% | 24040 |
| Case D. | 6.7534e-07 | 1.0463e-06 | 7.9512e-07 | 1.1974e-06 | 100% | 24080 |

[a] Success means the the solution is less than $1e-5$ in 100 independent simulations.

[b] No. of evaluation for objective function

It is true that enhancing the generation numbers of DE as case C is the best way to achieve higher efficiency and precision. However, as we can see that No. of evaluation for objective function should be much more. This might be some kind of "No Free Lunch".

It should be noticed too many points for evaluating fractional order chaos system the individual represents are not worth. Because the most time consumption parts in the whole proposed method are to resolve the candidate systems. In our simulations 200 is selected.

Some of these system are easy to solve. However when it comes with the some individuals with bad combinations of time-delays, parameters and fractional orders , the methods to resolve the fractional order chaos systems in Section Methods might not converge as shown in the simulations for system (11). Then the whole proposed method Algorithms 1 might get into endless loops. To avoid the endless loops, we introduce a forced strategy to assign all the NAN and infinite numbers in the output as 1. Because the objective function (3) to be optimized is bigger than 0, so this forced strategy for assignment is reasonable.

To achieve a fine balance between the performance of the proposed methods and having enough sample data for credibility, we take the number of the points as 50, 100, according the existing simulations[51–66]. And the simulations in section Results results show it is effective too.

Here we have to say that this work is only about the estimation of the objective function (3) for time-delayed fractional order chaos systems in non-Lyapunov way. It can be concluded that DE in Algorithms 1 can be change to other artificial intelligence methods easily. For the cases that some fractional order differential equations are unknown but with definite orders $q$ have been discussed in Reference [42].

The performance of the proposed method is sensitive to a series factors, such as the initial point for each fractional order chaos system, sample interval, number of points, and predefined intervals for the unknown time-delays, fraction orders and parameters. Actually, these also lead to the candidate system divergent. A good combination of these is not easy to get. Some mathematical formula to get a good combination not by so many simulations will be introduced in the future studies.

## 4. Conclusions

In conclusion, it has to be stated that proposed Algorithms 1 for time-delayed fractional order chaos systems' identification in a non-Lyapunov way is a promising direction.

The inversion mechanism put consists of numerical optimization problem with unknown fractional order differential equations to identify the chaotic systems with a novel functional extrema model as in Equation (5). Simulation results demonstrate the effectiveness and efficiency of the proposed methods with the mathematical model in Section 3. This is a novel Non–Lyaponov way for fractional order chaos' unknown time-delays, parameters and fractional orders.

In the future, we will do further researches for the cases that neither the tiem-delays, fractional orders nor some fractional order equations are known. That is, the objective function is chosen as following equation (12) in the novel mathematic model in Section 2. In another words, the objective will be changed into as equation (12).

$$(\tilde{q}, \tilde{f})^* = \arg\min_{(\tilde{q},\tilde{f})} F \tag{12}$$

**References**

[1] N. H. Sweilam, M. M. Khader, A. M. S. Mahdy, Numerical studies for fractional-order logistic differential equation with two different delays, Journal of Applied Mathematics 2012 (2012) 14.

[2] S. Bhalekar, V. Daftardar-Gejji, A predictor-corrector scheme for solving nonlinear delay differential equations of fractional order, Journal of Fractional Calculus and Applications 1 (5) (2011) 1–8.

[3] V. Daftardar-Gejji, S. Bhalekar, P. Gade, Dynamics of fractional-ordered chen system with delay, Pramana 79 (1) (2012) 61–69.

[4] L. Yuan, Q. Yang, C. Zeng, Chaos detection and parameter identification in fractional-order chaotic systems with delay, Nonlinear Dynamics 73 (1-2) (2013) 439–448.

[5] V. Celik, Y. Demir, Chaotic Fractional Order Delayed Cellular Neural Network, Springer Netherlands, 2010, Ch. 27, pp. 313–320.

[6] V. Celik, Y. Demir, Chaotic dynamics of the fractional order nonlinear system with time delay, Signal, Image and Video Processing (2013) 1–6.

[7] H. KOBER, On fractional integrals and derivatives, The Quarterly Journal of Mathematics os-11 (1) (1940) 193–211.

[8] K. Diethelm, An efficient parallel algorithm for the numerical solution of fractional differential equations, Fractional Calculus and Applied Analysis 14 (3) (2011) 475–490.

[9] A. A. Kilbas, H. M. Srivastava, J. J. Trujillo, Theory and Applications of Fractional Differential Equations, Vol. Volume 204 of North-Holland Mathematics Studies, North-Holland, Amsterdam, Netherlands: Elsevier, 2006.

[10] S. Samko, A. Kilbas, O. Marichev, Fractional Integrals and Derivatives: Theory and Applications, CRC, 1993.

[11] G. Si, Z. Sun, Y. Zhang, W. Chen, Projective synchronization of different fractional-order chaotic systems with non-identical orders, Nonlinear Analysis: Real World Applications 13 (4) (2012) 1761–1771.

[12] M. S. Tavazoei, M. Haeri, Chaotic attractors in incommensurate fractional order systems, Physica D: Nonlinear Phenomena 237 (20) (2008) 2628–2637.

[13] S. Fazzino, R. Caponetto, A semi-analytical method for the computation of the lyapunov exponents of fractional-order systems, Communications in Nonlinear Science and Numerical Simulation (0).
URL http://dx.doi.org/10.1016/j.cnsns.2012.06.013

[14] M. S. Tavazoei, M. Haeri, A necessary condition for double scroll attractor existence in fractional-order systems, Physics Letters A 367 (1?) (2007) 102–113.

[15] I. Grigorenko, E. Grigorenko, Chaotic dynamics of the fractional lorenz system, Phys. Rev. Lett. 91 (2003) 034101. doi:10.1103/PhysRevLett.91.034101.
URL http://link.aps.org/doi/10.1103/PhysRevLett.91.034101

[16] Y. Tang, X. Zhang, C. Hua, L. Li, Y. Yang, Parameter identification of commensurate fractional-order chaotic system via differential evolution, Physics Letters A 376 (4) (2012) 457–464.

[17] L.-G. Yuan, Q.-G. Yang, Parameter identification and synchronization of fractional-order chaotic systems, Communications in Nonlinear Science and Numerical Simulation 17 (1) (2012) 305–316.

[18] W. Deng, S. E, D. Sun, P. Wang, D. Zhang, W. Xu, Fabrication of vertical coupled polymer microring resonator, in: Y.-C. Chung, S. Xie (Eds.), ICO20: Optical Communication., Vol. 6025 of Proceedings of the SPIE, 2006, pp. 334–339.

[19] L. Song, J. Yang, S. Xu, Chaos synchronization for a class of nonlinear

oscillators with fractional order, Nonlinear Analysis: Theory, Methods and Applications 72 (5) (2010) 2326–2336.

[20] R. X. Zhang, Y. Yang, S. P. Yang, Adaptive synchronization of the fractional-order unified chaotic system, Wuli Xuebao/Acta Physica Sinica 58 (9) (2009) 6039–6044.

[21] Z. M. Odibat, N. Corson, M. A. Aziz-Alaoui, C. Bertelle, Synchronization of chaotic fractional-order systems via linear control, International Journal of Bifurcation and Chaos 20 (1) (2010) 81–97.

[22] Z. Odibat, A note on phase synchronization in coupled chaotic fractional order systems, Nonlinear Analysis: Real World Applications 13 (2) (2012) 779–789.

[23] X. Wu, D. Lai, H. Lu, Generalized synchronization of the fractional-order chaos in weighted complex dynamical networks with nonidentical nodes, Nonlinear Dynamics 69 (1-2) (2012) 667–683, export Date: 27 June 2012 Source: Scopus.

[24] E. Kaslik, S. Sivasundaram, Analytical and numerical methods for the stability analysis of linear fractional delay differential equations, Journal of Computational and Applied Mathematics 236 (16) (2012) 4027–4041.

[25] E. H. Doha, A. H. Bhrawy, S. S. Ezz-Eldien, A new jacobi operational matrix: An application for solving fractional differential equations, Applied Mathematical Modelling 36 (10) (2012) 4931–4943.

[26] H. Jian-Bing, X. Jian, D. Ling-Dong, Synchronizing improper fractional

chen chaotic system, J.Shanghai University (Natural Science Edition) 17 (6) (2011) 734–739.

[27] C. Li, G. Peng, Chaos in chen's system with a fractional order, Chaos, Solitons & Fractals 22 (2) (2004) 443–450.

[28] X. J. Wu, S. L. Shen, Chaos in the fractional-order lorenz system, International Journal of Computer Mathematics 86 (7) (2009) 1274–1282, cited By (since 1996): 5 Export Date: 27 June 2012 Source: Scopus.

[29] S. Bhalekar, V. Daftardar-Gejji, Fractional ordered liu system with time-delay, Communications in Nonlinear Science and Numerical Simulation 15 (8) (2010) 2178–2191.

[30] S. Bhalekar, V. Daftardar-Gejji, Synchronization of different fractional order chaotic systems using active control, Communications in Nonlinear Science and Numerical Simulation 15 (11) (2010) 3536–3546.

[31] S. K. Agrawal, M. Srivastava, S. Das, Synchronization of fractional order chaotic systems using active control method, Chaos, Solitons & Fractals 45 (6) (2012) 737–752.

[32] X. Wang, X. Zhang, C. Ma, Modified projective synchronization of fractional-order chaotic systems via active sliding mode control, Nonlinear Dynamics 69 (1-2) (2012) 511–517.

[33] R. L. Bagley, R. A. Calico, Fractional order state equations for the control of viscoelasticallydamped structures, Journal of Guidance, Control, and Dynamics 14 (2) (1991) 304–311.

[34] Z. M. ODIBAT, N. CORSON, M. A. AZIZ-ALAOUI, C. BERTELLE, Synchronization of chaotic fractional-order systems via linear control, International Journal of Bifurcation and Chaos 20 (01) (2010) 81–97.

[35] G. Si, Z. Sun, H. Zhang, Y. Zhang, Parameter estimation and topology identification of uncertain fractional order complex networks, Communications in Nonlinear Science and Numerical Simulation.
URL http://dx.doi.org/10.1016/j.cnsns.2012.05.005

[36] G. Si, Z. Sun, H. Zhang, Y. Zhang, Parameter estimation and topology identification of uncertain fractional order complex networks, Communications in Nonlinear Science and Numerical Simulation (0).
URL http://dx.doi.org/10.1016/j.cnsns.2012.05.005

[37] X. Wu, Z. Sun, F. Liang, C. Yu, Online estimation of unknown delays and parameters in uncertain time delayed dynamical complex networks via adaptive observer, Nonlinear Dynamics 73 (3) (2013) 1753–1768.

[38] Z. Sun, G. Si, F. Min, Y. Zhang, Adaptive modified function projective synchronization and parameter identification of uncertain hyperchaotic (chaotic) systems with identical or non-identical structures, Nonlinear Dynamics 68 (4) (2012) 471–486.

[39] F. Gao, F.-x. Fei, Q. Xu, Y.-f. Deng, Y.-b. Qi, Identification of unknown parameters and orders via cuckoo search oriented statistically by differential evolution for non-commensurate fractional order chaotic systems, eprint arXiv:1208.0049.

[40] I. Petráš, Fractional-Order Chaotic Systems, Vol. 0 of Nonlinear Physical Science, Springer Berlin Heidelberg, 2011, pp. 103–184.
URL http://dx.doi.org/10.1007/978-3-642-18101-6_5

[41] I. Petráš, Fractional calculus, Vol. 0 of Nonlinear Physical Science, Springer Berlin Heidelberg, 2011, pp. 7–42.
URL http://dx.doi.org/10.1007/978-3-642-18101-6_2

[42] G. Fei, F. Feng-xia, X. Qian, D. Yan-fang, Q. Yi-bo, I. Balasingham, Reconstruction mechanism with self-growing equations for hyper, improper and proper fractional chaotic systems through a novel non-lyapunov approach, Arxiv.

[43] W. H. Deng, C. P. Li, Chaos synchronization of the fractional lü system, Physica A: Statistical Mechanics and its Applications 353 (0) (2005) 61–72.

[44] R. Storn, K. Price, Differential evolution – a simple and efficient heuristic for global optimization over continuous spaces, J. of Global Optimization 11 (4) (1997) 341–359.

[45] F. Gao, H. Tong, Computing two linchpins of topological degree by a novel differential evolution algorithm, International Journal of Computational Intelligence and Applications 5 (3) (2005) 335–350.

[46] F. Gao, Computing unstable period orbits of discrete chaotic system though differential evolutionary algorithms basing on elite subspace, Xitong Gongcheng Lilun yu Shijian/System Engineering Theory and Practice 25 (2005) 96–102.

[47] C.-W. Chiang, W.-P. Lee, J.-S. Heh, A 2–opt based differential evolution for global optimization, Appl. Soft Comput. 10 (4) (2010) 1200–1207.

[48] P. Kenneth, M. S. Rainer, A. L. Jouni, Differential Evolution: A Practical Approach to Global Optimization (Natural Computing Series), Springer-Verlag New York, Inc., 2005.

[49] K. Diethelm, N. Ford, A. Freed, A predictor-corrector approach for the numerical solution of fractional differential equations, Nonlinear Dynamics 29 (1-4) (2002) 3–22.

[50] S. Bhalekar, V. Daftardar-Gejji, A predictor-corrector scheme for solving nonlinear delay differential equations of fractional order, Journal of Fractional Calculus and Applications, An International Journal 1 (5) (2011) 1–9.

[51] U. Parlitz, Estimating model parameters from time series by autosynchronization, Physical Review Letters 76 (8) (1996) 1232.

[52] L. Li, Y. Yang, H. Peng, X. Wang, Parameters identification of chaotic systems via chaotic ant swarm, Chaos, Solitons & Fractals 28 (5) (2006) 1204–1211.

[53] W.-D. Chang, Parameter identification of chen and lu systems: A differential evolution approach, Chaos Solitons & Fractals 32 (4) (2007) 1469–1476.

[54] F. Gao, X.-J. Lee, F.-X. Fei, H.-Q. Tong, Y.-B. Qi, Y.-F. Deng, I. Balasingham, H.-L. Zhao, Parameter identification for van der pol-

duffing oscillator by a novel artificial bee colony algorithm with differential evolution operators, Applied Mathematics and Computation DOI: 10.1016/j.amc.2013.07.053doi:10.1016/j.amc.2013.07.053.

[55] F. Gao, Z.-Q. Li, H.-Q. Tong, Parameters estimation online for lorenz system by a novel quantum-behaved particle swarm optimization, Chinese Physics B 17 (4) (2008) 1196–1201.

[56] F. Gao, H. Q. Tong, Parameter estimation for chaotic system based on particle swarm optimization, Acta Physica Sinica 55 (2) (2006) 577–582.

[57] F. Gao, H. Gao, Z. Li, H. Tong, J.-J. Lee, Detecting unstable periodic orbits of nonlinear mappings by a novel quantum-behaved particle swarm optimization non-lyapunov way, Chaos, Solitons & Fractals 42 (4) (2009) 2450–2463.

[58] F. Gao, J.-J. Lee, Z. Li, H. Tong, X. Lü, Parameter estimation for chaotic system with initial random noises by particle swarm optimization, Chaos, Solitons & Fractals 42 (2) (2009) 1286–1291.

[59] K. Yang, K. Maginu, H. Nomura, Parameters identification of chaotic systems by quantum-behaved particle swarm optimization, International Journal of Computer Mathematics 86 (12) (2009) 2225–2235.

[60] F. Gao, Y. Qi, I. Balasingham, Q. Yin, H. Gao, A novel non-lyapunov way for detecting uncertain parameters of chaos system with random noises, Expert Systems with Applications 39 (2) (2012) 1779–1783.

[61] F. Gao, Y. Qi, Q. Yin, J. Xiao, Solving problems in chaos control though

an differential evolution algorithm with region zooming, Applied Mechanics and Materials 110-116 (2012) 5048–5056.

[62] F. GAO, F.-X. FEI, Y.-F. DENG, Y.-B. QI, B. Ilangko, A novel non-lyapunov approach through artificial bee colony algorithm for detecting unstable periodic orbits with high orders, Expert Systems with Applications.
URL http://dx.doi.org/10.1016/j.eswa.2012.04.083

[63] F. Gao, Y. Qi, Q. Yin, J. Xiao, An novel optimal pid tuning and on–line tuning based on artificial bee colony algorithm, in: The 2010 International Conference on Computational Intelligence and Software Engineering (CiSE 2010), IEEE, Wuhan, China, 2010, pp. 425–428.

[64] F. Gao, Y. Qi, Q. Yin, J. Xiao, Online synchronization of uncertain chaotic systems by artificial bee colony algorithm in a non–lyapunov way, in: The 2010 International Conference on Computational Intelligence and Software Engineering (CiSE 2010), IEEE, Wuhan, China, 2010, pp. 1–4.

[65] F. Gao, Y. Qi, Q. Yin, J. Xiao, An artificial bee colony algorithm for unknown parameters and time–delays identification of chaotic systems, in: the Fifth International Conference on Computer Sciences and Convergence Information Technology (ICCIT10), IEEE, Seoul, Korea, 2010, pp. 659 – 664.

[66] F. Gao, Y. Qi, Q. Yin, J. Xiao, A novel non–lyapunov approach in discrete chaos system with rational fraction control by artificial bee colony

algorithm, in: 2010 International Conference on Progress in Informatics and Computing(PIC-2010), IEEE, Shanghai, China, 2010, pp. 317 – 320.